\documentclass[twocolumn]{aa}

\usepackage{graphicx}

\usepackage{txfonts}
\usepackage{natbib}

\newcommand\Tstrut{\rule{0pt}{2.6ex}}         
\newcommand\Bstrut{\rule[-0.9ex]{0pt}{0pt}}   

\begin{document}

\title{Estimates of the change rate of solar mass and gravitational constant based on the dynamics of the Solar System}
\titlerunning {Estimates of change rate of solar mass and gravitational constant}
\authorrunning{Pitjeva et al.}
         \author{E. V. Pitjeva \inst{1} \and N. P. Pitjev\inst{1,2} \and D. A. Pavlov \inst{1,3} \and C. C. Turygin\inst{1,2}}

   \institute{Institute of Applied Astronomy of Russian Academy of Sciences (IAA RAS),
Kutuzov Quay~10, 191187 St.Petersburg, Russia \\
          \email{evp@iaaras.ru}
          \and
St.Petersburg State University, Universitetski pr. 28, 198504 Petrodvoretz, Russia \\
\email{ai@astro.spbu.ru}
\and
         St. Petersburg Electrotechnical University,  ul. Professora Popova 5, 197376 St. Petersburg, Russia \\
          \email{dapavlov@etu.ru}
             }

   \date{Received 11 November 2020; accepted 2021}

  \abstract{The estimate of the change rate of the solar gravitational parameter
$\mathrm{d}(GM_{\odot})/\mathrm{d}t$ is obtained from  processing modern 
positional observations  of planets and spacecraft. Observations were processed
 and parameters were determined basing on the numerical planetary ephemeris 
EPM2019. The obtained annual decrease in
  solar mass $M_{\odot}$ accounts for the loss through radiation
  ${\dot M}_{{\odot}\mathrm{rad}}$, through the outgoing solar wind 
${\dot M}_{{\odot}{\mathrm{wind}}}$,
  and for the material falling on the \object{Sun} ${\dot M}_{{\odot}\mathrm{fall}}$.
  The estimated relative value is within
 $-13.4\cdot 10^{-14} < (\dot M_{\odot} /M_{\odot})_{\mathrm{rad}+\mathrm{wind}+
\mathrm{fall}} < -8.7\cdot 10^{-14}$ per year.
  The following range for the change rate of the gravitational constant $G$ was
 obtained: 
$-2.9 \cdot 10^{-14} < \dot G /G < + 4.6 \cdot 10^{-14}$  per year $(3\sigma$).
  The new result reduces the interval for the change in $G$ and narrows the 
limits of possible deviations for alternative gravitational theories from general relativity.}

   
   \maketitle

\section{Introduction}

The source of the \object{Sun}'s energy is thermonuclear fusion in its interior,
which means that the mass of the \object{Sun} must be changing. The generated energy $\Delta E_{\odot}$
obeys Einstein's formula $E = mc^2$ and thus gives
the corresponding decrease in solar mass $\Delta M_{\odot}$.
The other physical effect that causes the decrease in solar mass
was detected much later with the discovery of
the continuous flow of plasma out into the interplanetary space, called
solar wind \citep{Parker1958}. 

Coronal ejections add material to the solar wind. This is amplified during
periods of active \object{Sun}. Although the combined mass loss of the \object{Sun}
due to solar radiation and solar wind is millions of tons per second,
the relative change is very small. First experimental results on the
change in the product $GM_{\odot}$ (the gravitational parameter of
the \object{Sun}) were obtained in 2012~\citep{Pitjeva2012} from the motions
of the planets in the Solar System.

From the equations of motion of celestial mechanics
it is impossible to determine a direct change in mass of the central body $ \Delta M_ {\odot}$, and it is only possible
to determine the change products of $ \Delta(GM_ {\odot})$ because the equations
include products of mass by the gravitational constant $G$.
The important problem of constancy or variability of $G$ arises here.

The question of the possible variability of $G$,
an important fundamental physical quantity, has been discussed for a long time.
In the general theory of relativity (GRT), $G$ is a main
parameter and has a constant value. However, some works have considered a variable $G$:
\citet{Dirac1937, Dirac1938} and \citet{Milne1937}.
In versions of the modified theory of gravity, as well as in alternative theories of gravity,
$G$ can change and is in fact a parameter (see, e.g., reviews of \citet{Uzan2003, Uzan2011} and \citet{Chiba2011}
about how $GM_{\odot}$ affects the motion of the planets).
The alternative theories of gravity include scalar-tensor theories \citep{Fujii2003}, quantum-gravitational theories
\citep{Bonanno2004, Smolin2015}, models of superstring theory, and cosmology with the variable $G$ \citep{Hanimeli2020}.
It is important to verify the constancy or variability of  $G$  for cosmological
theories and for possible modifications in the theory of gravity. Data from various fields of astrophysics can be useful for this purpose.

Local (in time and space) constraints were obtained from data for
double pulsars \citep{Zhu2019}, for exoplanets \citep{Masuda2016},
and on primary nucleosynthesis \citep{Alvey2020}.
The range of opportunities for the experimental evaluation of theories
gradually expands, and the answer to this question is currently sought through
physical and astronomical experiments.
The strongest restrictions on variations of the gravitational constant $G$
are obtained today from the dynamics of the Solar System.
High-precision theories of motion have been designed and built for the celestial
bodies of the Solar System:
DE (Development ephemeris, \citealt{folkner2014planetary}),
EPM (Ephemerides of planets and the \object{Moon}, \citealt{pitjeva2019epm}),
and INPOP (Int\'egrateur Num\'erique Plan\'etaire de l'Observatoire de Paris,
\citealt{viswanathan2018new}).
These ephemerides are publicly available
\footnote{\url{https://ssd.jpl.nasa.gov/?planet.eph.export}}
\footnote{\url{http://iaaras.ru/dept/ephemeris/epm}}
\footnote{\url{https://www.imcce.fr/inpop}.}

All this opened up an opportunity to consider the Solar System as a
huge laboratory for testing a number of important principles of
fundamental physics for gravitating bodies. In particular, finding a
change for the gravitational parameter of the \object{Sun} $\Delta
(GM_{\odot})$ allows us to analyze possible bounds for $\Delta M_{\odot}$
and $\Delta G$; the latter is important for testing the constancy or
variability of the gravitational constant.

\section{Effects of change in $GM_{\odot}$ on planetary motion}

To a certain extent, the masses of all bodies in the Solar System
can change (through the fall of asteroids, loss of gas from the atmosphere, etc.),
but the \object{Sun} is currently the only body for which we can detect the effect of
mass loss on gravity bodies because the mass of the \object{Sun} is many
orders of magnitude greater than the masses of planets and
other bodies. With the achieved accuracy of modern planetary ephemerides,
the change rate in $GM_{\odot}$
can be detected by a small change in the rate of the orbital movement of the bodies.

Changes in the masses of other bodies, including Jupiter, affect the planetary
motion much less strongly. The current accuracy of observations and planetary
theories means that these changes still cannot be detected.

With an assumed slow isotropic mass loss of $M_{\odot}$ associated with
solar radiation and the solar wind, the angular momentums of the planets are
preserved, but their semimajor axes $a_i$ change. At the same time, the shape,
eccentricity, and position of the pericenter are preserved. Each orbit gradually
transforms, remaining similar to itself, and resembles a spiral.

Direct search for a resizing of orbits due to the time evolution of the
value of $GM$ for the central body (as was noted by \citet{Pitjeva2012})
is ineffective because over a time interval of several tens or
hundreds of years, the effect is too small.
It is difficult to overcome the complexity in practical measurements of very
small effects against the background of existing
objective errors for the parameters of planetary orbits, even for high-precision observations.  It is important that along with the change
in the semimajor axes $a_i$, the orbital periods $T_i$ change, however. This key point opens up the possibility of detecting
the effect of a small change in $GM_{\odot}$, as there is a shift in the
position of bodies in longitude $\Delta l_i$, and the longitudinal
displacement increases with each revolution of the planet around the \object{Sun}. The change rate of the movement leads to a shift in the position of the
body throughout the orbit in proportion to  the square of the time interval
$(\Delta t)^2$. Using  long-range observations
allows us to find the change over time $\mathrm{d}(GM_{\odot})/\mathrm{d}t$.

\section{Planetary ephemeris EPM2019 and observations}

The value of $\mathrm{d}(GM_{\odot})/\mathrm{d}t$ was determined as part of a least-squares
fitting of a large number of parameters to a large number of observations,
which is normally done to build planetary ephemerides, in this case, the ephemerides of the planets and the \object{Moon} (EPM). These ephemerides were 
created in the 1970s in support of Russian space flight missions; later, their
development continued in the Insitute of Applied Astronomy of the Russian
Academy of Sciences (IAA RAS). The EPM contain coordinates and
velocities of the \object{Sun}, the \object{Moon}, the eight major planets, \object{Pluto}, the three
largest asteroids (\object{Ceres}, \object{Pallas}, and \object{Vesta}) an four transneptunian objects (TNOs: \object{Eris}, \object{Haumea},
\object{Makemake}, and \object{Sedna}), as well as the lunar libration and the difference
between the Terrestrial time and the Barycentric dynamical time (TT$-$TDB).
The modern EPM cover more than 400 years (1787--2214)
except for the ``long'' ephemeris EPM2017H, which spans an interval of more than 13,100 years.
The dynamical model of the EPM is based on the parameterized post-Newtonian
N-body metric for general relativity in the barycentric coordinate
system (BCRS) and the TDB timescale. The motion of the \object{Sun}, the
planets (including \object{Pluto}), and the \object{Moon} (as point-masses) obeys the
Einstein-Infeld-Hoffmann relativistic equations. 

Standard procedures were carried out to determine corrections
  to the parameters values using the iterative least-squares method
  fed with the entire set of observations. The method was repeatedly
  applied until the corrections to the parameters became smaller than their error
  estimates. Several iterations were usually required. All
  observations were weighted according to their accuracy. The
  parameters depend on the observations used. For early radar
  observations of planetary surfaces, it is necessary to know the
  parameters of the planet surface topography; for the three
  Martian landers (Viking 1 and 2, and Pathfinder),
  13 parameters of the rotation of \object{Mars} and three parameters of
  the coordinates for each lander need to be determined. To process the ranging observations
  of planetary orbiters, factors of the solar plasma correction were
  determined for each conjunction of the planet with the \object{Sun} (21
  parameters). The most important parameters are the orbital
  elements of the nine planets (including \object{Pluto}) and 18 main satellites of
  the outer planets (162 parameters), however. A total of 264 parameters were
  determined for all observations.
  
The modern EPM is significantly more advanced than the
EPM2010, both for an improved dynamical model and for a
wider range of high-precision observations.  For the change
the solar mass, the accuracy of the mean planetary longitudes is most important because the semimajor axes of the planets and
the periods of the orbital motion are sensitive to the change in solar mass (see section 2). Compared to EPM2010, the errors of
the mean longitudes in EPM2017 have decreased for almost all planets,
except for \object{Neptune}, for which only one two-dimensional point
was obtained from the data of the Voyager 2 spacecraft. The errors in mean
longitudes of \object{Mercury} and \object{Mars} have decreased by more than an order of
magnitude based on observations of the Martian orbiters and the
MESSENGER (\object{Mercury} Surface, Space Environment, Geochemistry, and Ranging)
orbiter and taking the Lense-Thirring
relativistic effect into account \citep[Table 3]{Pitjeva2018masses}.

We determined the change in solar mass simultaneously with
all parameters, except for the masses of asteroids, which correlate
with each other and would affect the uncertainty of the change in solar
mass. The asteroids masses were determined independently
from $\mathrm{d}(GM_{\odot})/\mathrm{d}t$, but
together with the other parameters of the planetary ephemeris.

The time interval of the observations exceeds 100 years
(1913--2017) and includes optical observations and normal points of
radio-technical observations (see Table~\ref{table:1}).  The
  number of high-precision radar observations on which the EPM are
  based increases constantly. When EPM2010 was released,
  about 800,000 observations were collected. More than 850,000
  observations are used in the current version of the EPM2019 planetary
  theory. However, individual spacecraft
  measurements are not used in ephemerides, only 51858 normal points. 
 Normal points were made from raw observations by William
  Folkner (NASA JPL); each normal point represents the observations of
  one spacecraft revolution around the planet (these observations
  correlate with each other). Table 1 shows the number of normal
  points for ranging data and the number of observations for optical
  data. For the ephemerides of the inner planets, only high-precision
  radio-technical measurements are currently used, which cover a time
  interval of more than half a century. Optical observations are
  inferior in accuracy by several orders of magnitude, and were not
  used for these planets.

            

The most accurate are radio ranging measurements of planetary
orbiters:
\object{Mercury} (MESSENGER 2011--2015, 0.7 m), \object{Venus} (Venus Express 2006--2012, 3 m),
 \object{Mars} (Mars Express 2009--2015,1.5 m; Odyssey 2002--2017, 1 m; Mars Reconnaissance Orbiter 2006--2017, 1 m),
 \object{Jupiter} (Cassini 2004--2014, 20 m), and \object{Saturn} (Juno 2016--2017, 11 m). The values in parentheses are
the root-mean-square postfit residuals (see \citep{Pitjeva2018mass,Pitjeva2018masses,Pitjeva2019}
for details).

Recent optical observations of good $(<  1^{\prime\prime})$  
accuracy were obtained fom observatories: TMO (USA), Flagstaff
(USA), Lowell (USA), and Pico dos Dias (Brazil).
The dynamical model has become more accurate than the model in
EPM2010. The EPM2010 model included mutual perturbations from the major planets,
the \object{Sun}, the \object{Moon}, and 301 asteroids chosen because they strongly perturbed
\object{Mars} and \object{Earth}; it included the main relativistic perturbations;
perturbations due to the solar oblateness $J_2$; and perturbations from the
21 largest TNOs. The EPM2019 model in addition 
includes perturbations from the other nine largest TNOs, 
perturbations from discrete massive two-dimensional rotational rings of main-belt asteroids 
 and TNO rings, and perturbations of the next order of
smallness, which also affect the motion of the planets. It also includes
the relativistic Lense-Thirring effect and the \object{Jupiter} trojans
\citep{Pitjeva2018mass, Pitjeva2018masses, Pitjeva2019}.

 Since EPM2017, the EPM are being built using the ERA-8
software \citep{Pavlov2016}. Recent improvements to this software
relevant to planetary dynamics are the inclusion of two-dimensional
discrete rotating rings accounting for the Kuiper belt and for the
small asteroids in the main belt
\citep{Pitjeva2018mass,Pitjeva2018masses}, the relativistic
Lense-Thirring effect, and two numerous groups of Jupiter Trojans
\citep{Pitjeva2019}. The Lense-Thirring effect is especially important
for the determination of the time-varying value of $GM_{\odot}$
because it allows us to build a much more correct orbit of \object{Mercury}, which
is fit to the ranging observations of the MESSENGER
spacecraft. An important technical achievement was also made with
the new multistep integrator \citep{Aksim2020}, which is capable of
handling delay differential equations in a manner that does not
decrease the performance. Because the planets and the \object{Moon} are
integrated together and because the lunar equations contain delay
\citep{Pavlov2016}, the new integrator has allowed us to build the
ephemeris twice as fast.

The following value was estimated for the change rate of $(GM_{\odot})$:

$\frac{\mathrm{d}(GM_{\odot})/\mathrm{d}t}{GM_{\odot}} = (-10.2 \pm 1.4) \cdot 10^{-14}$ per year
$(3\sigma)$.

This is the value for the annual change in the solar gravitational parameter $(GM_ {\odot})$
and includes the annual change in solar mass $M_{\odot}$ and a possible annual change in the gravitational constant
$G$. We note the following relation for $\dot {(GM_{\odot})}$:
$$\frac{\mathrm{d}(GM_{\odot})/\mathrm{d}t}{GM_{\odot}} = \frac{\dot G}{G} + \frac{\dot M_{\odot}}{M_{\odot}}, $$
hence

$-11.6\cdot 10^{-14} < \dot G /G + \dot M_{\odot} /M_{\odot} < -8.8\cdot 10^{-14}$ per year.   $ (3\sigma)$

\section{Change rate in solar mass}

The change rate in solar mass 
($\dot M_{\odot}$) has several components.
First of all, the solar mass decreases as a result of thermonuclear fusion
with the release of energy  ${\dot M}_{{\odot}{\mathrm{rad}}}$ , which provides
continuous powerful radiation from the \object{Sun}.
Second, the plasma of the solar corona has a very high temperature and is emitted 
into interplanetary space in the form of an accelerated solar wind
${\dot M}_{{\odot}{\mathrm{wind}}}$, 
which continuously carries matter away, along with
charged wind particles and solar material.
Third, a certain amount of dust partices
falls onto the \object{Sun} due to the Poynting-Robertson effect;
also certain number of comets, asteroids, and meteoroids
fall onto the \object{Sun} due to the evolution of their orbits. We denote
the change rate due the falling material $({{\dot M}_{\odot}})_{\mathrm{fall}}$.
Considering these components, the total change rate  in solar mass is

$\dot M_{\odot} /M_{\odot} = (\dot M_{\odot} /M_{\odot})_{\mathrm{rad}} + (\dot M_{\odot} /M_{\odot})_{\mathrm{wind}}
+ (\dot M_{\odot} /M_{\odot})_{\mathrm{fall}}$.

\subsection{Solar radiation}

The value of the nominal solar radiation per second,
covering the full range of electromagnetic waves, and adopted by the International
Astronomical Union in 2015, is equal to

$L_{\odot} = 3.828 \cdot 10^{26} \, W = 3.828 \cdot 10^{33} \, \mathrm{erg/s}$

(Resolution B3, International Astronomical Union, 2015).

This radiation energy corresponds to
solar mass loss of 

$({{\dot M}_{\odot}})_\mathrm{rad} = -6.760\cdot 10^{-14} \, M_{\odot}$ per year.

There are small variations in the solar flux around this value. They are
related to the 11-year
 solar cycle with a maximum deviation of about 0.1\%, as well as to changes
of about 0.2\%, related to the 27-day period of the solar rotation around its axis \citep{Frohlich2004}.
These changes involve the ultraviolet, visible, and infrared regions of
the spectrum, with large changes at shorter wavelengths. Variations in solar
radiation are clearly traced
by solar magnetism. The active regions alter the local radiation,
and the contrasts depending on  the wavelength relative to the quiet \object{Sun}
determine the variability of the flow. The solar emissivity also reacts
to the subsurface convection and hot plasma flows in the \object{Sun}. On the shortest
timescales, the total radiation
shows five-minute fluctuations with an amplitude of $ \approx 0.003\%$ and
can increase to 0.015\% during the largest solar flares.

In general, fluctuations in solar flux during the solar cycle are small and
amount to $\sim 0.1-0.2\%$ \citep{Wu2018, Tagirov2019}.
Based on the maximum estimate for fluctuations, the average annual loss
of solar mass $\dot M_{{\odot}{\mathrm{rad}}}$ therefore is within the range of

$-6.801\cdot 10^{-14} <(\dot M_{\odot} /M_{\odot})_{\mathrm{rad}} < -6.719\cdot 10^{-14}$ per year $(3\sigma).$

The possible range for the deviation from the nominal average value
${\dot M}_{{\odot}_\mathrm{rad}}$ due to fluctuations associated with various solar activities 
is insignificant compared with the fluctuations 
for the mass that is carried away per unit time by
the solar wind $({{\dot M}_{\odot}})_\mathrm{wind}$. 

\subsection{Solar wind}

The solar wind is a stream of charged particles coming out of
the base of the solar corona. Material is added to this magnetized plasma stream,
emitted as a result of coronal emissions, the amount of which increases during
the period of active \object{Sun}. The stationary part of the flow consists of fast
(from 600 km/s to 800 km/s)
and slow (up to 450--500 km/s) wind \citep{Belcher1971}. These components
can vary with time and and fluctuate in density.
The fast solar wind occurs in coronal holes. 
The origins of the slow solar wind are not established
clearly yet and are still discussed 
\citep{Antonucci2005, Abbo2016}. They have been associated with active areas on the \object{Sun}. Generally, the solar wind is
a large-scale plasma outflow, and its mass is almost entirely composed
of protons and alpha particles.  The electron mass contributes approximately three orders of magnitude less.
The proportion of other elements is insignificant.
 Coronal emissions significantly affect fluctuations in the solar wind density.
 The number and intensity of the emissions depend on solar activity, the number
of sunspots, and on the solar cycle phase. The emissions of magnetized plasma
are usually closely related to solar flares \citep{Compagnino2017}.
The main share of the emissions falls on latitudes $-60^{\circ} \le b \le 60^{\circ}$.
The solar mass loss due to coronal mass ejections is an order
of magnitude lower than that of the solar wind \citep{Mishra2019}, however.

Additional information about the properties of the solar wind has  begun
to be obtained through the Parker Solar Probe spacecraft (PSP) made by NASA. It operates in the vicinity
of the \object{Sun} and gradually moves to a trajectory close to the \object{Sun}
\citep{Halekas2020, Chen2020, Rouillard2020}.

During periods of high solar activity, slow winds are observed at all
latitudes. At low solar activity the latitudinal structure is bimodal: the slow
wind is concentrated in near-equatorial zones, and the fast wind in near-polar zones.

To estimate the mass that is carried away with the solar wind, averaged over the solar
cycle, we used
data from the Ulysses spacecraft (NASA)\footnote{http://ufa.esac.esa.int/ufa/data}.
Ulysses operated from 1990 to 2008, and the data cover the entire
solar activity cycle 23, the final half of solar cycle 22, and the beginning of solar cycle 24.
It is important that the trajectory of the spacecraft was almost perpendicular
to the solar equator, while perihelion (1.35 AU) and aphelion (5.4 AU) were almost exactly in the
plane of the ecliptic.
The SWOOPS instrument (Solar Wind Observations Over the Poles of the Sun) measured the characteristics and densities of ion and
electron fluxes. From 1990 to 2008,
the device made about 
600,000 measurements every 15 minutes.
The temperature, density, and velocity of protons and alpha particles data
was registered.
When we processed the data, we used measurements averaged for each hour
of spacecraft work.

The Ulysses data clearly show the latitudinal distribution of the density and
velocity, correlated with the solar magnetic structure. Depending on
the phase of the solar cycle, the fast and slow winds are more noticeable or
fainter. When we found a stream of particles for the spacecraft position,
it was considered the same at this time for the entire latitude related
to the equator of the \object{Sun}.
The results were summarized for all latitudes during the entire operation of
the Ulysses spacecraft for about 18 years.

The mass carried away by the solar wind was estimated from the recorded
proton and alpha particle flux (SWOOPS). The plasma of the solar wind still contains
a stream of electrons, but the mass lost through electrons is more than three
orders of magnitude lower than the mass lost through high-mass particles.  The
fluctuations for the flow of protons and alpha particles are far lower.

The registration interval of particles by Ulysses began in solar cycle 22
and ended in cycle 24. It significantly exceeded
the average period of the solar cycle.
To obtain the average annual mass loss due to the solar wind
and take the variability during the solar cycle into account, it was averaged
using the entire operating interval of the spacecraft. The average period
for changes in solar activity was taken to be $\tau = 11.2$ years.
The average annual mass loss due to the solar wind is

$(\dot M_{\odot} /M_{\odot})_{\mathrm{wind}} = (-4.8 \pm 1.8)\cdot 10^{-14} \mathrm{yr}^{-1}
(3\sigma)$.

or

$-6.6\cdot 10^{-14} <(\dot M_{\odot} /M_{\odot})_{\mathrm{wind}} < -3.0\cdot 10^{-14} \mathrm{yr}^{-1} 
(3\sigma).$

An estimate of the mass carried away by the solar wind for a year was
found by processing data from the Ulysses spacecraft over the entire
18-year operation of the Ulysses. Averaging was carried out at each
11.2-year interval of the solar activity period with a shift of one year
from 1990 to 1997 to cover the entire observation interval up to 2008
in 11-year cycles. The value $\pm 1.8 \times 10^{-14}$ corresponds
to the random error found from variations in the average annual solar
mass loss due to the solar wind. About 67\% of the mass was carried
away with the fast wind, and about 33\% with the slow wind.

\subsection{Estimate of the mass falling onto the Sun}

The fall of small comets onto the \object{Sun} was recorded by spacecraft observing
the \object{Sun} and the solar corona. Interplanetary dust,
meteor matter and asteroids should also be considered part of a possible falling mass.
In this case,
we can only talk about an approximate estimate because there are not enough observational and experimental data.
Results of the PSP spacecraft can improve the situation.

\subsubsection{Dust of the interplanetary medium}

The dust environment in the Solar System is found from \object{Mercury} to the
Kuiper belt and might be
present in the Oort cloud. It is known from spacecraft observations that
it also includes interstellar dust particles passing through our Solar System
\citep{Grun1993, Baguhl1996, Altobelli2006}.
The proportion of the interstellar component is small, however, and is
three orders of magnitude lower than the density of interplanetary dust
belonging to the Solar System.

The interplanetary dust cloud is strongly concentrated toward the ecliptic
plane, and the dust particles  spent several thousand years in it, but
the interplanetary dust complex in the Solar System is constantly
replenished with dust lost by comets and asteroids.
Dust particles of the interplanetary medium interact with photons of solar
radiation, lose angular moment under the action of
the Poynting-Robertson effect, and gradually approach the \object{Sun} in a spiral \citep{Burns1979}. They are also exposed to solar pressure, thermal
heating, and sublimation, and when faced with the flow of the solar wind,
the dust experiences additional outward-directed pressure \citep{Klacka2012, Klacka2014}.
The lifetime of dust particles in the densest interplanetary environment extending
from the main asteroid belt toward the \object{Sun} is estimated to be
$\sim 10^3 $ years. There is constant exchange due to emission  
by comets and small fractions formed during asteroid collisions.

A population of very small, nanometer-sized dust particles is carried away by
the solar wind \citep{Juhasz2013}. According to the PSP
spacecraft, a large proportion of dust particles in the vicinity of the \object{Sun}
acquires hyperbolic velocities and is ejected from the Solar System \citep{Szalay2020}.

Thermal sublimation of dust particles predicts a dust-free zone $\sim$ 4--5 $R_{\odot}$.
The results of the PSP indicate the existence of a radiation mechanism
of dust destruction  \citep{Hoang2020}, which breaks large grains into very
small particles. They are in turn destroyed by the bombardment of protons from the solar
wind. As a result, the dust-free zone in the immediate vicinity of the \object{Sun} is likely
 $ \sim 8 R_{\odot}$.

The total mass of dust in the inner Solar System, including the main belt,
is low and is estimated to have a mass of about one asteroid with a
diameter of 20 to 30 km.
Dust particles drift and dust renewal takes place over several thousand years.
If all the dust moving toward
the \object{Sun} were to reach its surface, then the mass of dust falling onto the \object{Sun} due to the 
Poynting-Robertson effect would be about or lower than 
$+10^{-17} \, M_{\odot}$ per year. Solar radiation pressure, heating, and sublimation interfere with the approach to the surface of solar
dust particles, however, as does the pressure of the solar wind. The total value
$(\dot M_{\odot} /M_{\odot})_{dust}$ 
will therefore be far lower, and the contribution
 of the dust environment to the
material that falls onto the \object{Sun} is negligible.

\subsubsection{Comets and asteroids}
                      
The contribution of comets and asteroids to the mass falling onto the \object{Sun} may be more considerable
 than that of dust.
A substantial amount of data has been accumulated about objects that were detected
or passed in the vicinity of the \object{Sun}. Since 1996, permanent
observations from the space observatory Solar and Heliospheric Observatory (SOHO) have been made,
according to which 4000 comets were discovered by mid-June 2020,  most of them very small. 
Observations of close passages have been obtained from the space observatories: Solar Terrestrial Relations Observatory (STEREO A and B), the Solar Dynamics Observatory (SDO), and the PSP, which reached the solar vicinity in 2019.

Very few of the objects survive close proximity to the \object{Sun}. A close
passage of the \object{Sun} is usually accompanied by partial or severe destruction
of the body and the appearance of small fragments \citep{Sekanina2018}. \citet{Granvik2016} derived a critical perihelion value
$16 R_{\odot} = 0.074$ AU after
which active destruction of asteroids and comets begins to occur.

The perihelia of near-solar comets (Sun-grazing comets) are within
$3.5 R_{\odot}$ (this is inside the Roche limit for liquid bodies).
Several comets have been recorded whose
trajectories crossed the solar photosphere. Close passes led to the
destruction of the cores, sublimation, and ionization, to the loss of mass,
 momentum, and energy of comets in the solar corona. The cores of
most small comets are completely sublimated by solar radiation when they pass
in the vicinity of perihelion.
The properties of these small objects, mostly owned by the Kreutz Group,
have been analyzed (\citet{Biesecker2002, Marsden2005, Knight2010, Combi2019}).

Examples for complete destruction and disappearance of the core are
the near-solar comets C/2011 N3 (SOHO) \citep{Schrijver2012}
and C/2011 W3 (Lovejoy), which were observed by
the SDO \citep{Sekanina2012, McCauley2013}.
Comet C/2012 S1 ISON experienced significant destruction already on the approach
to perihelion \citep{Sekanina2015}.

The absence of meter-high fragments in the immediate vicinity of the \object{Sun},
according to the PSP observations \citep{Wiegert2020},
is explained by the fact that in the  process of destruction near the \object{Sun},
asteroids eventually gradually disintegrate to millimeter-sized particles.
The destruction of asteroids aids the concomitant erosion of material
under the effect of high-speed particles and meteoroids near the \object{Sun}.
It is difficult for small particles to reach the solar surface because 
the formed fragments and other small fractions near the \object{Sun} experience
heating, sublimation, and  solar pressure, leading the detritus and ejected
material of comets and asteroids away from the \object{Sun}.
A small amount of comet and asteroid matter does impact the \object{Sun}.

When small solid bodies (asteroids) approach the \object{Sun}, strong
radiation and heat effects give off gas molecules, dust particles, and
larger regolith particles \citep{Delbo2014}.
At small heliocentric distance, for example, for comets near
the \object{Sun}, the dust quickly dissociates and the gas becomes ionized
\citep{Povich2003, Bryans2012}, and the entire object
is often destroyed \citep{Biesecker2002}.

Observations thus indicate that only a small fraction of the substance can
reach the solar surface. The rarity of events and
the smallness of the bodies themselves indicates
that the estimate of the value
for the mass falling onto the \object{Sun} in our previous works has been significantly overstated.
This means that the estimate of the mass falling from comets and asteroids should be
significantly reduced
compared to previous values \citep{Pitjeva2012, Pitjeva2013}.
The top estimate of the increase in solar mass due to the falling matter of
comets in the previous model, as was immediately noted
in the paper \citep{Pitjeva2012}, was highly overestimated,

$(\dot M_{\odot} /M_{\odot})_{\mathrm{fall}} < +3.2\cdot 10^{-14}$ per year.

Our new top estimate for the average amount of mass of material falling onto the
 \object{Sun} is

$(\dot M_{\odot} /M_{\odot})_{\mathrm{fall}} < +1.0\cdot 10^{-14 }$ per year,

\noindent which is reduced by about three times and probably still overestimated by an order of 
magnitude or more.

\section {Limits of possible change in gravitational constant $G$}

For further estimates, we used the upper and lower bounds  
of the change rate of solar mass  $M_{\odot}$ 
  due to to radiation and solar wind within
a $3\sigma$ error, taking the estimate of the fall of comets and asteroids into account,
 $$(\dot M_{\odot} /M_{\odot})_{\mathrm{rad}+\mathrm{wind}+\mathrm{fall}} = 
(\dot M_{\odot} /M_{\odot})_{\mathrm{rad}} + (\dot M_{\odot} /M_{\odot})_{\mathrm{wind}}
 + (\dot M_{\odot} /M_{\odot})_{\mathrm{fall}}.$$

Taking the estimates for each component of the change rate in solar mass 
into account, we obtain
    
$-13.4\cdot 10^{-14} < (\dot M_{\odot}/M_{\odot} )_{\mathrm{rad}+\mathrm{wind}+\mathrm{fall}} 
< -8.7\cdot 10^{-14}$ per year.

For the change rate of the solar gravitational parameter $GM_{\odot}$, we obtained

$-11.6\cdot 10^{-14} < \dot G/G + \dot M_{\odot}/M_{\odot} < -8.8\cdot 10^{-14} $ per 
year ($3\sigma$).

From these two inequalities we find a range for a possible change rate  in the gravitational constant $G$ per year:

$-2.9\cdot 10^{-14} < \dot G /G < +4.6\cdot 10^{-14}$ year$^{-1}$ ($3\sigma$).

The new result, obtained with EPM2019 planetary solution, refines the ranges of the
possible change rate for the gravitational constant $G$ as compared to the
previous estimate \citep{Pitjeva2012}, obtained with EPM2010.  The estimates of
the possible interval for the change in $G$  obtained with different methods and
authors in the last eight years are shown in Table~\ref{table:1}. The last line
shows the result obtained in this work.

\begin{table*}
   \caption{Previous estimates of the change in $G$}
   \label{table:1}
   \centering
   \begin{tabular}{l c l}
    \hline
    \hline
   Authors \Tstrut\Bstrut & $s=\dot G /G$ year$^{-1}$ & Comment \\
   \hline \\
   \citet{Pitjeva2012}   & $-4.2\cdot 10^{-14} < s <+7.5\cdot 10^{-14}$ & planetary dynamics \\
   \citet{Fienga2015}    & $|s|<8\cdot 10^{-14}$ & planetary dynamics \\ 
    \citet{Hofmann2018}   & $s=(7.1 \pm 7.6)\cdot 10^{-14}$ & lunar laser ranging (LLR) \\
   \citet{Genova2018}    & $|s|<4\cdot 10^{-14}$ & Mercury's MESSENGER mission \\
   \citet{Bellinger2019} & $s=(2.1 \pm  2.9)\cdot 10^{-12}$ & asteroseismology \\
\citet{Bonanno2020} & $|s|<2\cdot 10^{-13}$ & helioseismology \\   
This paper &  $-2.9\cdot 10^{-14} < s <+4.6\cdot 10^{-14}$ & planetary dynamics \\   \hline
   \end{tabular}
\end{table*}

\section { Conclusion}

The estimate of the change rate for the solar gravitational parameter  $\dot {(GM_{\odot})}$ is obtained from processing 
modern positional observations of planets and spacecraft.
The observations were processed and the parameters determined based on a new version of the planetary ephemeris EPM2019 developed in the IAA RAS.

The new planetary ephemeris EPM2019 was constructed using observational data for planets
and spacecraft and a refined dynamical model of the Solar System.
The dynamical model uses new mass estimates for large bodies of the main asteroid
belt and a total mass estimate for small asteroids, their debris,
and dust using a new discrete model.
A similar refinement was made for the Kuiper belt.
The dynamical model includes the combined effect of two numerous and compact
groups of Jupiter Trojans at L4 and L5.
In EPM2019, the relativistic Lense-Thirring effect,
especially significant for \object{Mercury} and \object{Venus}, was taken into account.

The observational data include high-precision measurements that are important
for this work. They have been obtained relatively recently with the MESSENGER (\object{Mercury})
and Juno (\object{Jupiter}) spacecraft.
A new estimate of the change in solar gravitational parameter
$GM_{\odot}$ was found.
Taking into account the estimate for the loss of mass by the \object{Sun} averaged
over the solar cycle, new restrictions on the change in gravitational
constant $G$ from above and below were found.
The current annual decrease in solar mass $M_{\odot}$, taking into account
the mass lost $({\dot M}_{\odot})_\mathrm{rad}$ (solar radiation) and
 $({\dot M}_{\odot})_\mathrm{wind}$ (solar wind),
and the mass $({\dot M}_{\odot})_\mathrm{fall}$ of the material falling onto the \object{Sun},
is estimated to lie within
$$
-13.4\cdot 10^{-14} < (\dot M_{\odot} /M_{\odot})_{\mathrm{rad}+\mathrm{wind}+\mathrm{fall}}
< -8.7\cdot 10^{-14}\ \text{per year.}
$$

For the change in gravitational constant $G$, the following constraint was obtained:
$$
-2.9 \cdot 10^{-14} < \dot G /G < + 4.6 \cdot 10^{-14}\ \text{per year}\ (3\sigma).
$$

The new result reduces the interval for changing $G$ and narrows the possible limits
of deviation from general relativity for alternative gravitational theories.

\begin{acknowledgements} The authors thank Dan Aksim for his help with the LaTeX and BibTeX processing of the manuscript.
\end{acknowledgements}

\bibpunct{(}{)}{;}{a}{}{,}
\bibliographystyle{aa} 
\bibliography{references} 

\end{document}